# A METHOD FOR RAPID DETERMINATION OF OIL AND WATER CONTENT IN GEOLOGICAL FORMATIONS

D. Vartsky[1,*], M. B. Goldberg[2], V. Dangendorf[3], I. Israelashvili[1,5], I. Mor[4], D. Bar[4], K. Tittelmeier[3], M. Weierganz[3], B. Bromberger[3], A. Breskin[1]

[1] *Weizmann Institute of Science, Rehovot, 76100, Israel*
[2] *Herzbergstr. 20, 63584 Gründau, Germany*
[3] *Physikalisch-Technische Bundesanstalt (PTB), 38116 Braunschweig, Germany*
[4] *Soreq NRC, Yavne 81800, Israel*
[5] *Nuclear Research Center of the Negev, P.O.Box 9001, Beer Sheva, Israel*

**ABSTRACT**
A novel method utilizing Fast Neutron Resonance Transmission Radiography is proposed for rapid, non-destructive and quantitative determination of the weight fractions of oil and water in cores taken from subterranean or underwater geological formations. Its ability to distinguish water from oil stems from the unambiguously-specific energy-dependence of the neutron cross-sections for the principal elemental constituents. Furthermore, the fluid weight fractions permit determining core porosity and oil and water saturations. In this article we show results of experimental determination of oil and water weight fractions in 10 cm thick samples of Berea Sandstone and Indiana Limestone formations, followed by calculation of their porosity and fluid saturations.
The technique may ultimately permit rapid, accurate and non-destructive evaluation of relevant petro-physical properties in thick intact cores. It is suitable for all types of formations including tight shales, clays and oil sands.

**INTRODUCTION**
Routine oil-drilling core analysis consists of measuring porosity, permeability, and fluid or gas saturation [1,2]. Most prevalent analysis techniques are based on destructive analysis of small plug samples removed from the core. More recent techniques include X-ray CT [3] and MRI analysis [4,5] that could, in principle, be applied non-destructively to larger core samples.
Nuclear geophysics is a discipline that assists oil, gas and uranium exploration, both in nuclear borehole-logging and analysis of core samples [6]. Middleton et al, [7], investigated thermal neutron radiography to estimate the rock porosity and relative fluid saturation in 5 mm-thick rock slices. The use of thermal neutrons does not permit distinguishing between water and oil, because it relies mainly on the attenuation of hydrogen. De Beer et al [8] also used thermal-neutron radiography to provide internal structure images of rocks, in order to determine the effective porosity of the object. Nshimirimana et al [9] examined the precision of porosity calculations in 14-17 mm thick rock samples using thermal neutron radiography. Lanza et al, [10] investigated thermal neutron computerized tomography to image the distribution of hydrogenous liquids (oil or water) in a 25.4 mm-diameter core. As in the above-mentioned studies it cannot distinguish between oil and water either. In certain cases deuterated water is introduced into the porous media, in order to study immiscible fluid flow by thermal neutron tomography [11]. A recent review [12] of thermal-neutron imaging of hydrogen-rich fluids in geo-materials discusses the non-destructive visualization of such fluids within diverse porous media.

---

*Corresponding author
e-mail: david.vartsky@weizmann.ac.il

In our previous work [19] we described a proof-of-principle study of the Fast Neutron Resonance Transmission (FNRT) radiography method for core analysis using synthetic samples of sand saturated with oil or water. In this paper we further evaluate the technique using thick, real formation cores.

**FAST NEUTRON RESONANCE TRANSMISSION RADIOGRAPHY**

A description of FNRT radiography has been given in [13-18] and its specific application for core analysis was detailed by Vartsky et al.[19]. Briefly, FNRT radiography is a method that exploits characteristics (resonances) in the neutron attenuation of the analysed constituents in order to determine the identity and proportions of substances within an object. A typical neutron energy-range is 1-10 MeV.

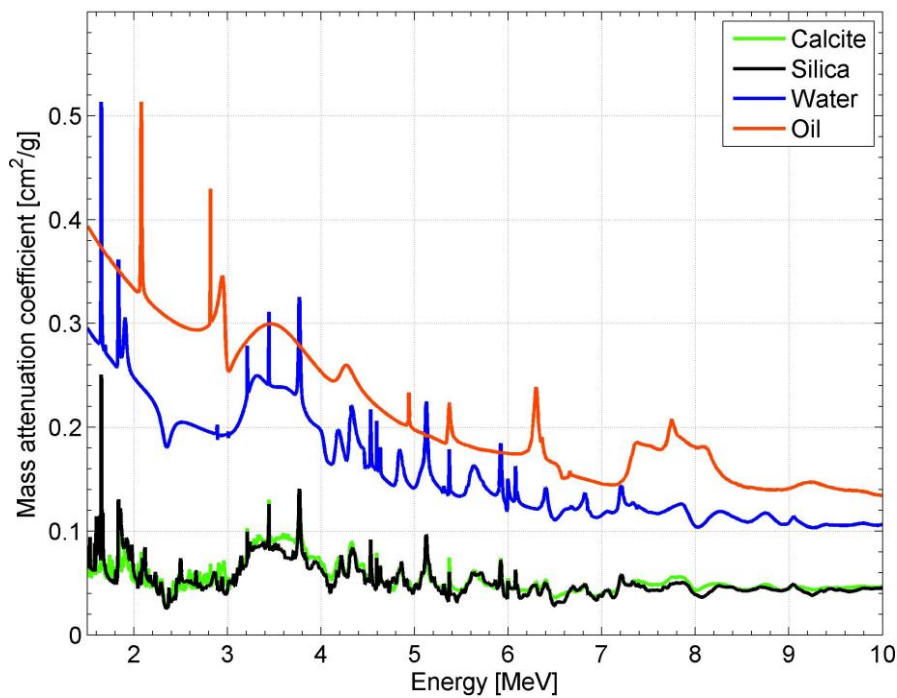

*Fig. 1 Mass-attenuation coefficients of silica, calcite, oil and water vs. neutron energy.*

Fig. 1 shows the energy dependence of the mass attenuation coefficients of calcite, silica, (the principal constituents of limestone and sandstone rocks respectively), oil and water. The values were calculated for the above substances using compiled neutron cross-sections [20] of their elemental constituents. It can be observed that the attenuation coefficients of the four substances exhibit different characteristic behaviour with neutron energy. This is due to resonances in the neutron interaction with the most abundant elements in materials, such as carbon in oil, oxygen in water, oxygen and silicon in silica and calcium, oxygen and carbon in calcite. In most elements the resonances occur mainly at lower neutron energies (below 8 MeV) and are due to compound nucleus formation. In such reaction the probability of interaction increases, when the energy of the incident neutron corresponds to an excited nuclear state of the resulting compound nucleus. Following this stage a neutron may be reemitted leading to elastic or inelastic resonance scattering. Hydrogen, present in oil and water does not exhibit any resonances in its attenuation coefficient, which decreases smoothly with neutron energy. Thus, for example, the resonant features in water are all due to resonances in cross-section of oxygen, which ride on a smooth hydrogen cross-section curve.

In FNRT the inspected object is irradiated with a broad spectrum of neutrons in the above-mentioned energy range. Dependent on the nature of the inspected object the transmitted neutron spectrum will exhibit dips and peaks at specific energies-thus carrying information about the object's composition. This is similar to observing characteristic absorption lines observed in other analytical spectroscopic techniques; for example atomic absorption method.

Fig. 2 schematically shows the FNRT irradiation configuration. An intact core within its protective sleeve is subjected to a broad-energy neutron beam in the energy range 1-10 MeV. The transmitted neutron spectrum is detected by a fast-neutron position-sensitive detector to provide mm-resolution imaging capability. In addition to position resolution the detector must be spectroscopic, i.e. it should provide information on the energy of the detected neutrons.
The shape of the core can be arbitrary and the method can provide the relevant information regardless of its geometry.

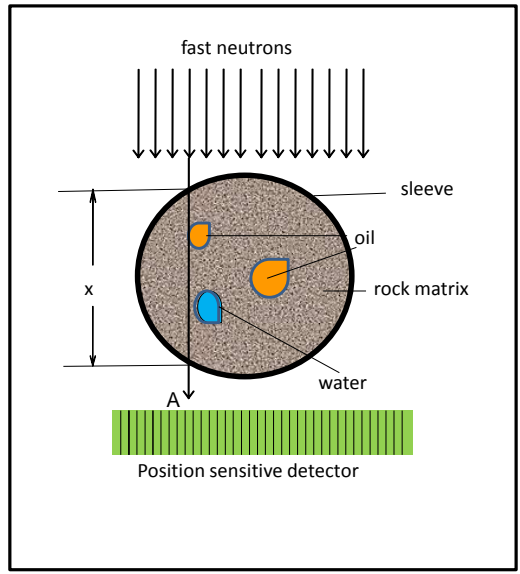

*Fig. 2 Schematic description of FNRT irradiation configuration of a core in its protective sleeve. The arrow marked "A" stands for part of the beam of fast neutrons that traverses the thickness x of the sample and impinges on a specific pixel within the array.*

If we assume that the inspected object, such as an oil-drilling core, consists mainly of porous rock matrix (eg. sandstone), oil and water; (we can ignore the presence of few mg/g of Cl which may be part of high salinity water, since for MeV energy neutrons its cross section is only few barns), the ratio $R_i$ of the transmitted-to-incident neutron flux at an energy **i** and at the position indicated in the drawing by the arrow (A) is:

$$R_i = \exp[-(\boldsymbol{\mu}_i^s \rho_s x + \boldsymbol{\mu}_i^o \rho_o x + \boldsymbol{\mu}_i^w \rho_w x)] \quad \text{Eq.1}$$

Where $\boldsymbol{\mu}_i^s$, $\boldsymbol{\mu}_i^o$, $\boldsymbol{\mu}_i^w$ and $\rho_s x$, $\rho_o x$, $\rho_w x$ are the mass attenuation coefficients and areal densities of dry formation (sandstone, limestone), oil, and water, respectively (see also Fig. 1). The densities $\rho$ in Eq. 1 are not the intrinsic physical densities of the substances, they represent the mean densities averaged over the trajectory $x$.
Since the spectrum may consist of **n** discrete neutron-energies, one can write **n** such equations. By taking a natural logarithm of $R_i$ one obtains a set of **n** linear equations where the areal densities are the unknowns of interest. This is an over-determined system, in which there are **n** linear equations with three unknowns (of these **n**, not all have the same sensitivity: in other words, depending on the element in question, the effective number of equations may be considerably smaller than the nominal **n**. Such a problem can be solved by a least-squares solution with bootstrapping or a Bayesian minimization method [21-23].

Once a solution for the 3 areal densities is found for a given detector pixel, we can determine the areal-density-ratio of oil or water to that of the dry rock. This yields the local weight-fractions of oil and water **$f_o$** and **$f_w$** in the traversed core, independent of sample thickness or shape. It must be noted that the fluid weight-fractions in the sample are determined independently, thus the oil-to-rock weight-ratio is independent of water content.

One can now display the map of oil or water weight-fractions for each individual pixel. Alternatively, by multiplying each pixel areal density by a pixel area we obtain the mass of each component in a volume defined by pixel area and height x and by integrating over all pixels obtain the total weight of oil, water and dry rock in the entire core, from which the average weight-fractions of oil and water **$F_o$** and **$F_w$** in the core can be determined regardless of the object shape, thickness or fluid distribution.

Prior to analysis of the core of interest we must calibrate our system using substances of known composition and physical densities. To this end we must determine experimentally the values of the mass attenuation coefficients vs. neutron energy for pure dry-rock of known grain-density, oil and water. This calibration procedure is necessary since there could be significant differences between rock and oil types from one drilling site to another.

Alternatively, or if such standards are unavailable, one may use calibrated elemental standards, such as Si, O, C, H, Ca Al and Mg to measure their mass-attenuation coefficients. In such a case, solving Eq. 1 will yield elemental areal densities, from which one can deduce the content of oil and water in the core.

From the determined oil and water average weight fractions $F_o$ and $F_w$ it is further possible to calculate the dry weight of the core (**$DW_C$**), the average rock porosity (**$\Phi$**) and average oil and water saturation levels ($S_o$, $S_w$) of the analysed core, provided we can measure the total weight-(**$TW_c$**) and volume ($V_c$) of the analysed core and that the grain-density of the rock, as well as the densities oil and water (**$\rho_G, \rho_o, \rho_w$**) are known, using the following equations:

$$\mathbf{DW_c} = \frac{TW_c}{(1+F_o+F_w)} \qquad \text{Eq. 2}$$

$$\mathbf{\Phi} = 1 - \frac{TW_c/V_c}{(1+F_o+F_w)\cdot\rho_G} \qquad \text{Eq. 3}$$

$$\mathbf{S_o} = \frac{F_o\cdot(TW_c/\rho_o)}{V_c\cdot(1+F_o+F_w)-(TW_c/\rho_G)} \qquad \text{Eq. 4}$$

$$\mathbf{S_W} = \frac{F_W\cdot(TW_c/\rho_W)}{V_c\cdot(1+F_o+F_w)-(TW_c/\rho_G)} \qquad \text{Eq. 5}$$

**EXPERIMENTAL PROCEDURES**

Preparation of Formation Samples

Evaluation of the technique was performed using rock formation samples of known properties. Three Berea Sandstone and three Indiana Limestone cubical samples 10x10x10 cm$^3$ in dimensions were prepared by Kocurek Company [24]. After cutting to the desired dimensions the samples were dried at 82°C in core-drying oven. Their weight was recorded. Two samples of each formation type were then inserted into a vacuum chamber and evacuated for approximately 1 hour. Following this step, the given fluid was pulled into the sample from the bottom at a very slow rate using a vacuum pump. After saturation the samples were weighed, the direction of "top" and "bottom" was marked and they were sealed in a container immersed in the saturation fluid for shipping. The saturation fluids were water and Odorless Mineral Spirit (OMS, density=0.748 g/cc) supplied by the Univar Company.

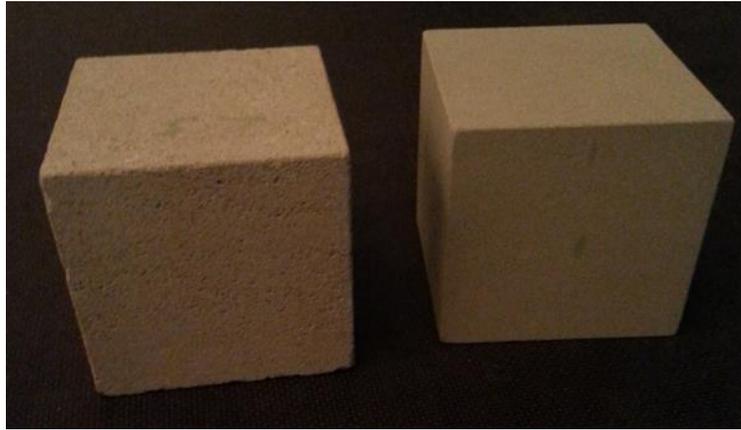

*Fig. 3 Indiana Limestone (left) and Berea Sandstone (right) samples*

Fig. 3 shows the dry limestone and sandstone samples and Table 1 summarizes the weights of all samples. The bulk volume of the samples was measured to be 1000±2.3 cc. Based on the weights, volumes and literature values of grain densities of sandstone and limestone (2.66 and 2.71 g/cc respectively) [25] we calculated the sample porosity and oil and water saturation levels, also shown in Table 1.

*Table 1 Weights, porosity and fluid saturation values of the analyzed samples*

| Sample No | Sample type | Weight (g) | Porosity (%) | Saturation(%) |
|---|---|---|---|---|
| 1 | Dry limestone (LS) | 2257.7±0. | 16.8±0.23 | --- |
| 2 | LS+oil | 2380.9±0. | * | 97.9±0.23 |
| 3 | LS+water | 2403.3±0. | * | 86.6±0.23 |
| 4 | Dry sandstone (SS) | 2112.3±0. | 20.6±0.23 | --- |
| 5 | SS+oil | 2270.0±0. | * | 102.4±0.23 |
| 6 | SS+water | 2323.7±0. | * | 102.7±0.23 |

*Assumed to be the same as for dry sample

The higher than 100% saturation values for sandstone samples could result from inaccurate grain density taken from literature. Porosity calculation and the resulting saturation values are rather sensitive to variations in grain density. A change in grain density by less than 1% can result in porosity change of about 3%.

For determining the mass-attenuation coefficients $\mu_i^s$, $\mu_i^o$, $\mu_i^w$ we used the dry samples of limestone and sandstone, pure OMS liquid supplied by Univar and regular tap water.

Neutron Irradiation Procedure

The experiment was performed using the CV28 isochronous cyclotron at Physikalisch-Technische Bundesanstalt (PTB), Braunschweig, Germany. Neutrons were produced by a 12 MeV deuterium beam impinging on a 3 mm thick Be target. The useful part of the neutron energy spectrum ranges from ca. 1 MeV up to 10 MeV [26]. Neutron spectroscopy was performed by the time-of-flight (TOF) method. In this method the time the neutron travels over a known distance between the target and the detector is measured and is converted to neutron energy. The deuteron beam was pulsed at a pulse repetition rate of 2 MHz and a pulse width of 1.7 ns. Average beam current was approximately 2 μA.

Neutrons were detected using a cylindrical 25.4 mm diameter x25.4 mm long liquid scintillator detector (NE213 type) positioned at 1247 cm from the target.

The analyzed samples were positioned between target and detector at a distance of 245 cm from the latter. The angle subtended by the detector was 0.058°, thus the diameter of the inspected region in the sample was 2 cm. The measurement time per sample ranged from 100-1000s.

**RESULTS AND DISCUSSION**

Uniformity Tests

As the samples were saturated with the fluids by pulling the liquid from the sample bottom, it was important to determine the uniformity of the fluid distribution along the direction of saturation. For this purpose the samples were scanned with the neutron beam directed perpendicularly to the direction of saturation from the bottom to the top of the sample in steps of 1 cm. The scans indicated that the fluids were uniformly distributed along the saturation direction to within ±2%.

As the samples were uniform in dimensions and composition there was no need to perform a high resolution radiographic scan and transmission measurements at a single point were performed using the liquid scintillator detector mentioned above.

Neutron Transmission Spectra

All transmission measurements were performed using the time-of flight (TOF) spectroscopy. In such measurements it is common to present the spectra vs. TOF rather than converting them to neutron energy. Fig. 4 shows the TOF spectra of the transmitted neutrons through a dry limestone and through water and oil-saturated limestone cores. As can be observed the spectrum is dominated by the shape of the dry limestone spectrum (the dominant absorber), nevertheless the proportions of various features are different for each configuration.

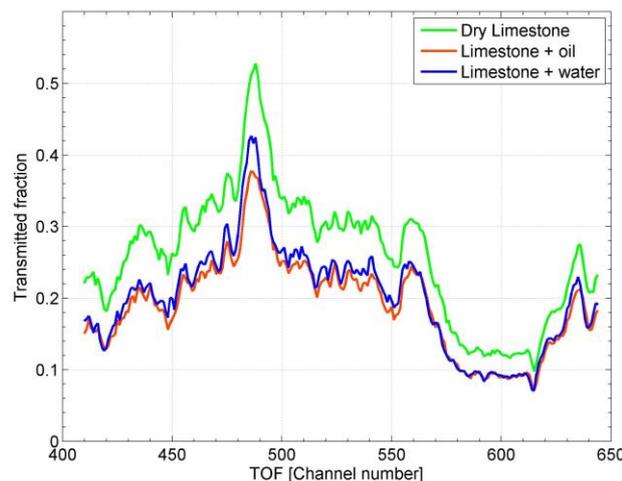

*Fig. 4 Transmission spectra vs TOF (expressed in channels) through 10 cm thick samples of dry limestone, limestone+oil and limestone+water. TOF range corresponds to neutron energy range of 1.7-4.2 MeV*

Fig. 5 shows the experimentally determined mass-attenuation coefficients of limestone, sandstone, oil (OMS) and water vs neutron TOF. The coefficients were determined by measuring neutron transmission through calibrated samples of pure dry limestone, dry sandstone, water and oil. Due to limited energy resolution of our experimental system the resonances are substantially broader and less pronounced than those based on compiled values of Fig. 1. These experimentally determined mass-attenuations are used as $\mu_i^s$, $\mu_i^o$, $\mu_i^w$ values in Eq. 1 for reconstructing the areal densities of dry rock, oil and water in the fluid saturated core samples. For reconstruction we used only a limited range of neutron TOF's corresponding to 1.7 to 4.2 MeV energy range. We found that this neutron energy range resulted in the best reconstruction.

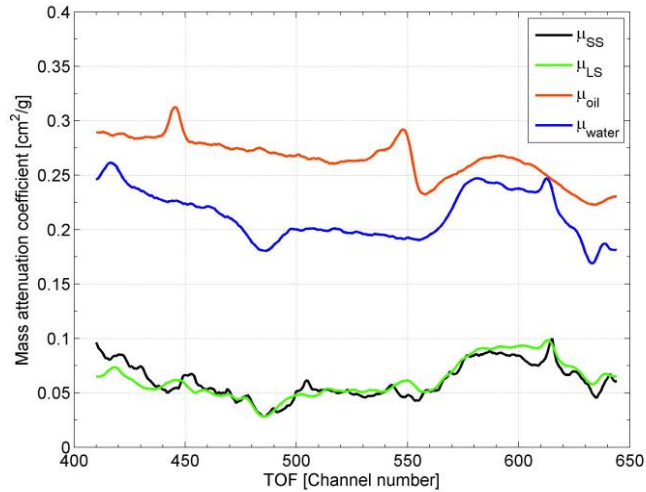

*Fig. 5 Experimentally determined mass attenuation coefficients for dry limestone, sandstone, water and oil (OMS).*

Reconstruction of Areal Densities

We used two methods for reconstructing the areal densities: 1) least-squares solution with bootstrapping and 2) WinBUGS program (Bayesian Inference Using Gibbs Sampling). [25]. Both types of analysis gave very similar reconstruction results and uncertainties. The reconstruction methods provide a probability distribution of the areal density for each constituent, indicating whether it is likely to be found in the inspected sample and what is the most probable areal density.

Fig. 6 shows the reconstructed experimental areal density distributions of dry sandstone core, oil and water in sandstone+oil (left column) and sandstone+water (right column). Here we used the least squares method with bootstrapping. The mean and standard deviation of the distributions are also indicated. The distributions for limestone samples are similar in shape.

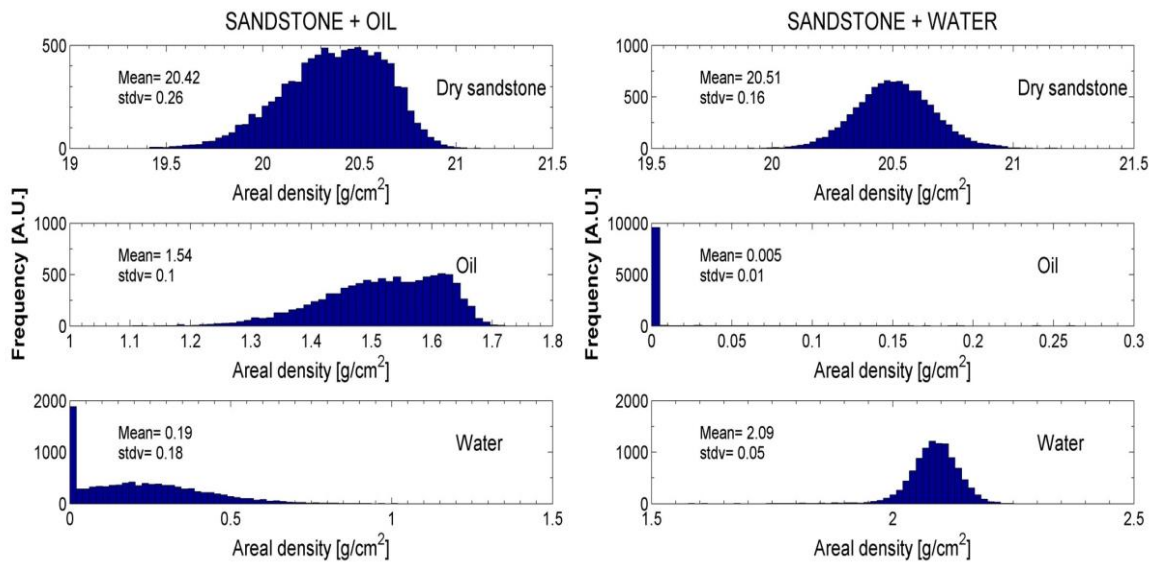

*Fig. 6 Distribution of least squares-reconstructed areal density of dry sandstone, oil and water in sandstone saturated with oil (left) and sandstone saturated with water (right).*

The ratio of reconstructed areal density of oil (or water) to that of dry core yields the weight fraction of each fluid in the core. Fig. 7 shows the experimentally determined oil *Fo* and water *Fw* weight fractions (in %) in each sample.

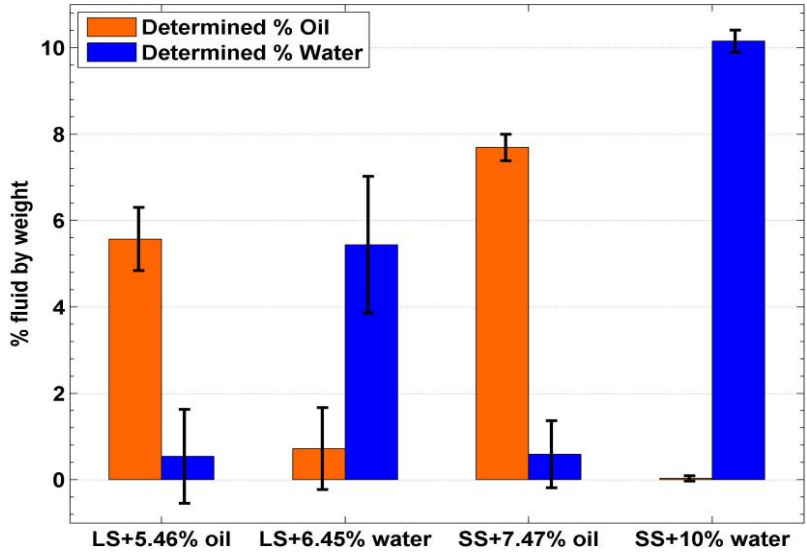

*Fig. 7 Experimentally determined weight percentage of oil (orange) and water (blue) in LS+5.46% oil, LS+6.45% water, SS+7.47% oil and SS+10% water*

Table 2 summarizes all experimental results: % weight of oil $F_o$ and water $F_w$, dry core weight-DWc, porosity-Φ, oil and water saturations-$S_o$;$S_w$ calculated using equations 2-5. The expected values determined by the prior weight and volume measurements are shown in square brackets.

*Table 2 Experimental $F_o$ and $F_w$ and calculated dry weight, porosity and saturations*

| Case | $F_o$ [%] | $F_w$ [%] | DWc [g] | Φ [%] | $S_o$[%] | $S_w$[%] |
|---|---|---|---|---|---|---|
| LS+oil | 5.57±0.7[5.46] | 0.54±1.1[0] | 2243±27[2257] | 17.3±0.2[16.8] | 96.4±12.7[98] | 7±14[0] |
| LS+water | 0.72±0.9[0] | 5.44±1.6[6.45] | 2264±39[2257] | 16.6±0.2[16.8] | 13±17[0] | 74.3±22[87] |
| SS+oil | 7.54±0.3[7.47] | 0.9±0.8[0] | 2093±18[2112] | 21.3±0.1 [20.3] | 98.9±4.2[102] | 9.1±8.6[0] |
| SS+water | 0.030±0.1[0] | 10.2±0.3[10] | 2108±5[2112] | 20.7±0.04 [20.3] | 0.36±0.86[0] | 103±2.6[103] |

[expected value]

The experimentally determined dry weights and porosity of the samples agree quite well with the expected values. The uncertainties in saturation values in limestone core case are relatively large 13-30% and can be mainly attributed to the errors in $F_o$ and $F_w$. The F and saturation values for fluids not present in the core are consistent with zero within their standard deviation. In addition, the shape of their frequency distributions (Fig. 6) is consistent with a characteristic distribution shape of substances which are not likely to be present in the inspected sample (strongly asymmetric distribution with a large value at zero).
Our previous Monte-Carlo calculations [19] indicated that the fluid content can be determined with sufficiently high accuracy and precision by irradiating a 10 cm thick core with a broad-energy neutron spectrum with no more than $10^6$ neutrons. However, the experimentally-calibrated reference values of the mass-attenuation coefficients of the standards (dry rock, oil and water) used in the reconstruction need to be determined with much higher precision than we have done so far, requiring at least ten-fold higher counting-statistics for the FNRTR spectra of these standards. Hence, although we collected about $10^6$ neutrons for the fluid saturated rocks in the present work, the uncertainties of the reconstructed experimental values, especially for limestone, appear to be higher than expected. We are confident that this is the sole procedural hurdle that the method still needs to surmount.

**CONCLUSIONS**
We describe a method based on Fast-Neutron Resonance Transmission (FNRT) radiography for a non-destructive, specific and quantitative determination of oil and water content in core samples. The application of fast neutrons can be useful in screening bulky objects such as thick rock cores, for which alternative probes, such as slow and epithermal neutrons, as well as low-energy X-rays, not only do not distinguish between hydrocarbons and water, but also suffer from limited penetration.

The method measures the average fluid/dry-core-weight ratio in the path traversed by the fast neutrons regardless of object shape, thickness or distribution. In principle the entire length of an intact core, within its protective sleeve can be scanned along the core length, providing information about the content distribution. The fluid weight-fractions in the interrogated sample are determined independently, thus the ratio of oil-to-rock weights is independent of the water content.

The measurement time is dependent on the incident neutron flux. Our measurement times/sample were about 1000s. The measurement time can be substantially reduced by using stronger neutron sources, and analysing many cores simultaneously using a large pixelated detector. We estimate that an operational facility will be able to analyse a core within minutes.

The experimentally determined fluid weight fractions F were determined with uncertainties of about 13-30%. We attribute these relatively large errors to insufficient counting statistics for our standards. The weight fraction values for fluids not present in the core are consistent with zero within their standard deviation. In addition, the shape of their frequency distributions (Fig. 6) is consistent with a characteristic distribution shape of substances which are not likely to be present in the inspected sample (strongly asymmetric distribution with a large value at zero).

We have demonstrated that if the total weight and volume of the core is available one can use the measured fluid weight fractions for the determination of the global core porosity and oil/water saturation. The experimentally determined dry weights and porosity of the samples agree quite well with the expected values. The uncertainties in saturation values in limestone core case are relatively large and can be mainly attributed to the errors in $F_o$ and $F_w$.

The FNRT method permits determining the fluid weight-fractions in any type of cores including tight shales, clays and oil sands. The method is also applicable for fluid-content evaluation in drill cuttings held in containers or bags.